# A New Perspective on Multi-user Power Control Games in Interference Channels

Yi Su and Mihaela van der Schaar

ABSTRACT

This paper considers the problem of how to allocate power among competing users sharing a frequency-selective interference channel. We model the interaction between selfish users as a non-cooperative game. As opposed to the existing iterative water-filling algorithm that studies the myopic users, this paper studies how a foresighted user, who knows the channel state information and response strategies of its competing users, should optimize its transmission strategy. To characterize this multi-user interaction, the Stackelberg equilibrium is introduced, and the existence of this equilibrium for the investigated non-cooperative game is shown. We analyze this interaction in more detail using a simple two-user example, where the foresighted user determines its transmission strategy by solving as a bi-level program which allows him to account for the myopic user's response. It is analytically shown that a foresighted user can improve its performance, if it has the necessary information about its competitors. Since the optimal solution of Stackelberg equilibrium is computationally prohibitive, we propose a practical low-complexity approach based on Lagrangian duality theory. Numerical simulations verify the performance improvements. Possible ways to acquire the required information and to extend the formulation to more than two users are also discussed.

## I. INTRODUCTION

The multi-user power control problem in frequency-selective interference channels was investigated from the game-theoretic perspective in several prior works, including [1]-[6]. In these multi-user wideband power control games, users are modeled as players having individual goals and strategies. They are competing or cooperating with each other until they agree on an acceptable resource allocation outcome. Existing research can be categorized into two types, *non-cooperative* games and *cooperative* games.

First, the formulation of the multi-user wideband power control problem as a non-cooperative game has appeared in several recent works [1][2]. An iterative water-filling (IW) algorithm was proposed to mitigate the mutual interference and optimize the performance without the need for a central controller [1]. At every decision stage, selfish users deploying this algorithm try to maximize their achievable rates by water-filling across the whole frequency band until a Nash equilibrium is reached. Alternatively, self-enforcing protocols are studied in the non-cooperative scenario, in which incentive compatible allocations are guaranteed [2]. By imposing punishments in the case of misbehavior and enforcing users to cooperate, efficient, fair, and incentive compatible spectrum sharing is shown to be possible.

Second, there also have been a number of related works studying dynamic spectrum management (DSM) in the setting of cooperative games [3]-[6]. Two (near-) optimal but centralized DSM algorithms, the Optimal Spectrum Balancing (OSB) algorithm and the Iterative Spectrum Balancing (ISB) algorithm, were proposed to solve the problem of maximization of a weighted rate-sum across all users [4][5]. OSB has an exponential complexity in the number of users. ISB only has a quadratic complexity in the number of users because it implements the optimization in an iterative fashion. An autonomous spectrum balancing (ASB) technique is proposed to achieve near-optimal performance autonomously, without real-time explicit information exchanges [6]. These works focus on cooperative games, because it is well-known that the IW algorithm may lead to Pareto-inefficient solutions [7], i.e. selfishness is detrimental in the interference channel.

In short, previous research mainly concentrates on studying the existence and performance of Nash equilibrium in non-cooperative games and developing efficient algorithms to approach the Pareto boundary in cooperative games. However, an important intrinsic dimension of this decentralized multi-user interaction still remains unexplored. Prior research does not consider the users' availability of information about other users and their potential to improve their performance when having this information. Hence, determining what is the best response strategy of a selfish user if it has the information about how the competing users respond to interference still needs to be determined. Moreover, it still needs to be established if such strategies can lead to a better performance than adopting the IW algorithm. It is important to look at these scenarios in order to assess the significance of information availability in terms of its impact on the users' performance in non-cooperative games, and show why selfish users have incentives to learn their environment and adapt their rational response strategies [8]. Intuitively, a "clever" user with more information in this non-cooperative game should be able to gain additional benefits [9].

Throughout this paper, we differentiate two types of selfish users based on their response strategies:

1) *Myopic user*: A user that always acts to maximize its immediate achievable rate. It is myopic in the sense that it treats other users' actions as fixed, ignores the dependence between its competitors' actions and its own action, and determines its response such that maximize its immediate payoff.

2) *Foresighted user*: A user that selects its transmission action by considering the long-term impacts on its performance. It anticipates how the others will react, and maximizes its performance by considering their reactions. It should be highlighted that additional information is required to assist the foresighted user in its decision making.

As opposed to previous approaches considering myopic users [1], we discuss in this paper how foresighted users should behave in non-cooperative power control games. We explicitly show that a strategic user can gain more benefit if it takes its competitors' information and response strategies into account. The concept of Stackelberg equilibrium is adopted in order to characterize the optimal power control strategy of a foresighted user by considering the response of its competing users. For the two-user case, we formulate the foresighted user's decision making to be a bi-level programming problem, show that the optimal solution is computationally prohibitive, and provide a low-complexity algorithm based on Lagrangian duality theory.

We also note that there are already some papers applying Stackelberg equilibrium to allocate the resources in networking [10]. However, the problems and the proposed solutions in these papers are completely different from this paper. The focus here is to study the strategic behavior of selfish users, which has not been yet investigated in multi-user interference channels.

The rest of the paper is organized as follows. Section II presents the non-cooperative game model and introduces the concept of Stackelberg equilibrium. In Section III, using a simple two-user example, we formulate the foresighted user's optimal decision making as a bi-level programming problem and discuss the computational complexity of its optimal solution. Section IV proposes a low-complexity dual-based approach and provides the simulation results. Section IV also discusses how the required information can be obtained by the strategic users and the problem formulation in general multi-user case. Conclusions are drawn in Section V.

## II. SYSTEM MODEL

In this section, we describe the mathematical model of the frequency-selective interference channel and formulate the non-cooperative multi-user power control game. We introduce the concept of Stackelberg equilibrium and prove the existence of this equilibrium in the power control game.

### A. System Description

Fig. 1 illustrates a frequency-selective Gaussian interference channel model. There are $K$ transmitters and $K$ receivers in the system. Each transmitter and receiver pair can be viewed as a player (or user). The whole frequency band is divided into $N$ frequency bins. In frequency bin $f$, the channel gain from transmitter $i$ to receiver $j$ is denoted as $H_{ij}^{f}$, where $f = 1, 2, \cdots, N$. Similarly, denote the noise power spectral density (PSD) that receiver $k$ experiences as $\sigma_k^f$ and player $k$'s transmit PSD as $P_k^f$. For user $k$, the transmit PSD is subject to its power constraint:

$$\sum\nolimits_{f=1}^{N} P_k^f \le \mathbf{P}_\mathbf{k}^{\mathbf{max}} . \tag{1}$$

Define $\boldsymbol{P}_k = \left\{P_k^1, P_k^2, \cdots, P_k^N\right\}$ as user $k$'s power allocation pattern. For a fixed $\boldsymbol{P}_k$, if treating interference as noise, user $k$ can achieve the following data rate:

$$R_k = \sum_{f=1}^{N} \log_2 \left( 1 + \frac{P_k^f \left|H_{kk}^f\right|^2}{\sigma_k^f + \sum\nolimits_{j \ne k} P_j^f \left|H_{jk}^f\right|^2} \right). \tag{2}$$

To fully capture the performance tradeoff in the system, the concept of a rate region is defined as

$$\mathcal{R} = \left\{ \left(R_1, \cdots, R_K\right) : \exists \left(\boldsymbol{P}_1, \cdots, \boldsymbol{P}_K\right) \ satisfying \ (1) \ and \ (2) \right\}. \tag{3}$$

Due to the non-convexity in the capacity expression as a function of power allocations, the computational complexity of optimal solutions (e.g., doing exhaustive search) in finding the rate region is prohibitively high. Existing works [4]-[6] aim to compute the Pareto boundary of this rate region and provide (near-) optimal performance with moderate complexity. Moreover, it is noted that cooperation among users is indispensable for this multi-user system to operate at the Pareto boundary. On the other hand, the interference channel can also be modeled as a non-cooperative game among multiple competing users. Instead of solving the optimization problem globally, the IW algorithm models the users as myopic decision makers [1]. This means that they optimize their transmit PSD by water-filling and compete to increase their transmission data rates with the sole objective of maximizing their own performance regardless of the coupling among users. Under a wide range of realistic channel conditions [1][13], the existence and uniqueness of the competitive optimal point (Nash equilibrium) is demonstrated and it can be obtained by the IW algorithm, which significantly outperforms the static spectrum management algorithms.

Throughout this paper, we also concentrate on the non-cooperative game setting. In the IW algorithm, users are assumed to be myopic, i.e., they update actions shortsightedly without considering the long-term impacts of taking these actions. We argue that the myopic behavior can be further improved because it neglects the coupling nature of players' actions and payoffs. In contrast with previous approaches, we study the problem of how a foresighted user should behave rather than taking myopic actions. This investigation provides us some insights to the following question: why should a strategic user sense its environment and learn the response strategies of its competitors and consequently, what is the benefit that a foresighted user can achieve compared with the myopic case?

To illustrate the foresighted behavior, Fig. 2 shows a simple Stackelberg game [11]. Note that in this game, the row player has a strictly dominant strategy [12], *Down*. Therefore, two players will end up with a (*Down*, *Left*) play if the row player is myopic. However, if the row player is aware of the column player's coupled reaction, they will end up with a (*Up*, *Right*) play, which leads to an increased payoff for both players. It is worth noticing that additional information is needed to attain this performance improvement. The row player needs to know the payoff and the response strategy of the column player. To formulate how a strategic user can take foresighted actions, we introduce the concept of Stackelberg equilibrium. The next subsection will define the Stackelberg equilibrium and show its existence in the power control game.

*B. Stackelberg Equilibrium*

Let $\mathcal{G} = \left[\mathcal{K}, \{\mathcal{A}_k\}, U_k\right]$ represent a game where $\mathcal{K} = \{1, \cdots, K\}$ is the set of players, $\mathcal{A}_k$ is the set of actions available to user $k$, and $U_k$ is the user $k$'s payoff [12]. In the power control game, user $k$'s payoff $U_k$ is the its achievable data rate $R_k$ and its action set $\mathcal{A}_k$ is the set of transmit PSDs satisfying constraint (1). Recall that the Nash equilibrium is defined to be any $\left(a_1^*, \cdots, a_K^*\right)$ satisfying

$$U_k\left(a_k^*, a_{-k}^*\right) \geq U_k\left(a_k, a_{-k}^*\right) \text{ } for \text{ } all \text{ } a_k \in \mathcal{A}_k \text{ } and \text{ } k = 1, \cdots, K \text{ }, \tag{4}$$

where $a_{-k}^* = \left(a_1^*, \cdots, a_{k-1}^*, a_{k+1}^*, \cdots, a_K^*\right)$ [12].

We also define the action $a_k^*$ to be a best response (BR) to actions $a_{-k}$ if

$$U_k\left(a_k^*, a_{-k}\right) \geq U_k\left(a_k, a_{-k}\right), \forall \text{ } a_k \in \mathcal{A}_k \text{ }. \tag{5}$$

The set of user $k$'s best response to $a_{-k}$ is denoted as $BR_k\left(a_{-k}\right)$.

The Stackelberg equilibrium is a solution concept originally defined for the cases where a hierarchy of actions exists between users [12]. Only one player is the *leader* and the other ones are *followers*. The leader begins the game by announcing its action. Then, the followers react to the leader's action. The Stackelberg equilibrium prescribes an optimal strategy for the leader if its followers always react by playing their Nash equilibrium strategies in the smaller sub-game. For example, in a two player game, where user 1 is the leader and user 2 is the follower, an action $a_1^*$ is the Stackelberg equilibrium strategy for user 1 if

$$U_1\left(a_1^*, BR_2\left(a_1^*\right)\right) \geq U_1\left(a_1, BR_2\left(a_1\right)\right), \forall a_1 \in \mathcal{A}_1 \text{ }. \tag{6}$$

For example, in Fig. 2, *Up* is the Stackelberg equilibrium strategy for the row player.

Next, we define Stackelberg equilibrium in the general case. Let $NE\left(a_k\right)$ be the Nash equilibrium strategy of the remaining players if player $k$ chooses to play $a_k$, i.e. $NE\left(a_k\right) = a_{-k}, \forall a_i = BR_i\left(a_{-i}\right), a_i \in \mathcal{A}_i, i \neq k$. The strategy profile $\left(a_k^*, NE\left(a_k^*\right)\right)$ is a Stackelberg equilibrium with user $k$ leading iff

$$U_k\left(a_k^*, NE\left(a_k^*\right)\right) \geq U_k\left(a_k, NE\left(a_k\right)\right), \forall a_k \in \mathcal{A}_k \text{ }. \tag{7}$$

If multiple Nash equilibria exist in the followers' sub-game, the definition of Stackelberg equilibrium becomes more complicated. Interested readers can refer to [10][14] for more details. This paper does not consider this case and focus on the channels where a unique Nash equilibrium exists in the sub-game [13]. In particular, the considered channel conditions are specified as follows:

***Considered Channels*** : Define $\alpha_{ij}^f = \left|H_{ij}^f\right|^2 / \left|H_{jj}^f\right|^2$ . For each frequency bin, we consider a $K \times K$ *channel gain matrix* $\mathrm{A}^f$ $(f = 1, \cdots, N)$, where $[\mathrm{A}^f]_{ij} = \alpha_{ij}^f$ for $i \neq j$ and $[\mathrm{A}^f]_{ii} = 0$ . This paper considers diagonally dominant channels in which $\left\|\mathrm{A}^f\right\|_2 < 1$ for any $f \in \{1, \cdots, N\}$. It is shown that a unique Nash equilibrium for the power control game exists in these channels and it can be achieved using the IW algorithm [13].

In fact, the requirement of hierarchic actions in the original definition of Stackelberg equilibrium can be removed in our problem if we consider the repeated interaction among all the users. Regardless of the initial action order, the foresighted user can always perform the Stackelberg strategy. As long as it changes its transmit PSD, the other myopic users will water-fill with respect to their updated noise-plus-interference PSDs to gain an immediate increase in transmission rates until the system converges to an equilibrium. We are interested in the performance achieved at the steady state. Therefore, the initial action order between the foresighted user and the myopic users does not influence the final outcome of this game. Note that initially we assume that a single foresighted user exists in this game. How the users should decide to play foresightedly or myopically and the extension to the cases where there are multiple foresighted users will be discussed in Section IV. The following theorem establishes the existence of Stackelberg equilibrium in the considered power control game.

***Theorem 1*:** Under the considered channel conditions, the Stackelberg equilibrium always exists in the multi-user power control game.

***Proof* :** Suppose user 1 is the only foresighted user in this game. First, user 1's maximal achievable rate in an interference-free environment is

$$R_1^{\max} = \sum_{f=1}^{N} \log_2 \left(1 + P_1^{f*} \left|H_{11}^f\right|^2 / \sigma_1^f\right), \tag{8}$$

where $P_1^{f*} = \left(\lambda - \sigma_1^f / \left|H_{11}^f\right|^2\right)^+$ is the water-filling solution, $(x)^+ = \max(0, x)$, and $\lambda$ is a constant satisfying the

constraint in (1) with equality.

Second, it has been shown that in the considered channels, the existence and uniqueness of Nash equilibrium are always guaranteed [13]. In the interference channel consisting of the $K-1$ followers, whatever form of $P_1^f \in \mathcal{A}_1$ user 1 chooses, they will regard user 1's transmit PSD as part of the fixed background noise PSD, i.e. $\tilde{\sigma}_j^f = \sigma_j^f + \left|H_{j1}^f\right|^2 P_1^f,\ j \neq 1$. Since the channel gains in the followers' sub-game still satisfy the sufficient condition in [13], the convergence to a unique Nash equilibrium always holds, i.e. a single $NE(a_1)$ exists for $\forall a_1 \in \mathcal{A}_1$.

To summarize, since $R_1$ is bounded, and for $\forall a_1 \in \mathcal{A}_1$, the remaining players' action will always lead to a Nash equilibrium, we have

$$0 \leq U_1\left(a_1, NE\left(a_1\right)\right) \leq R_1^{\max},\ \forall a_1 \in \mathcal{A}_1. \tag{9}$$

Therefore, there exist $a_1^* \in \mathcal{A}_1$ such that $U_1\left(a_1^*, NE\left(a_1^*\right)\right) = \sup_{a_1 \in \mathcal{A}_1}\left\{U_1\left(a_1, NE\left(a_1\right)\right)\right\}$. We can conclude that Stackelberg equilibrium always exists for this power control game. ■

## III. PROBLEM FORMULATION

In this section, we study how to achieve the Stackelberg equilibrium in the two-user case, and formulate the foresighted behavior as a bi-level programming problem. We analyze the computational complexity of the optimal solution, and show that the optimum is computationally intractable for the bi-level program. We start from the simplest two-user version, because it is illustrative for understanding the interactions emerging among competing users. The extension to the multi-user case will be discussed in Section IV.

### *A. A Bi-level Programming Formulation*

The Stackelberg equilibrium applied to the two-user power control game can be represented by a bi-level mathematical problem [14], in which the foresighted user acts as the leader and the other user behaves as the follower. The leader chooses a transmit PSD to maximize its own benefits by considering the response of its follower, who reacts to the leader's transmit PSD by water-filling over the entire frequency band. Hence, the

Stackelberg equilibrium can be found by solving the following optimization problem:

$$\begin{array}{ll} \begin{matrix} upper-level \\ problem \end{matrix} & \left\{ \begin{array}{ll} \max\limits_{\boldsymbol{P}_1} \sum\limits_{f=1}^{N} \log_2\left(1+\dfrac{P_1^f}{N_1^f+\alpha_2^f P_2^f}\right) & (a) \\ s.t.\ \sum_{f=1}^{N} P_1^f \le \mathrm{P}_{\mathbf{1}}^{\mathbf{max}},\ P_1^f \ge 0, & (b) \end{array} \right. \\ \begin{matrix} lower-level \\ problem \end{matrix} & \left\{ \begin{array}{ll} \boldsymbol{P}_2 = \arg\max\limits_{\boldsymbol{P}_2'} \sum\limits_{f=1}^{N} \log_2\left(1+\dfrac{P_2'^f}{N_2^f+\alpha_1^f P_1^f}\right) & (c) \\ s.t.\ P_2'^f \ge 0, \sum_{f=1}^{N} P_2'^f \le \mathrm{P}_{\mathbf{2}}^{\mathbf{max}}. & (d) \end{array} \right. \end{array} \quad (10)$$

where $N_1^f = \sigma_1^f \Big/ \left|H_{11}^f\right|^2, \alpha_1^f = \left|H_{12}^f\right|^2 \Big/ \left|H_{22}^f\right|^2, N_2^f = \sigma_2^f \Big/ \left|H_{22}^f\right|^2, \alpha_2^f = \left|H_{21}^f\right|^2 \Big/ \left|H_{11}^f\right|^2$. The sub-problem in (10.a)-(10.b) is called the *upper-level problem* and (10.c)-(10.d) corresponds to the *lower-level problem*. Recall that additional information is indispensable to formulate this bi-level program. This information includes the other user's channel condition $N_2^f$ and $\alpha_2^f$, maximum power constraint $\mathrm{P}_{\mathbf{2}}^{\mathbf{max}}$, and its response strategy, i.e. the IW algorithm. By letting $\boldsymbol{P}_1$ and $\boldsymbol{P}_2$ to be the transmit PSDs of the IW algorithm $\boldsymbol{P}_1^{NE}$ and $\boldsymbol{P}_2^{NE}$, we can see that the Nash equilibrium actually gives a lower bound of the problem in (10). Furthermore, by including the opponent's reaction into the lower-level problem, the user can avoid the myopic IW approach and potentially improve its performance. In addition, as we will show later, user 1's foresightedness turns out to even improve the myopic user's performance. Now we make several illustrative remarks by showing two simple examples.

***Remark 1*:** The Nash equilibrium achieved by the IW algorithm may not solve the bi-level program (10). In other words, there exist other feasible power allocation schemes that can attain strictly better performance than that of the Nash equilibrium.

***Example 1*:** We consider a two-user system with the parameters $N=2$, $N_1^1 = N_2^2 = 4$, $N_1^2 = N_2^1 = 1$, $\alpha_i^f = 0.5$ for $\forall i, f$, $\mathbf{P}_{\mathbf{1}}^{\mathbf{max}} = \mathbf{P}_{\mathbf{2}}^{\mathbf{max}} = 10$. In this simple two-channel scenario, it is easy to derive that $R_1 = \log_2\left[1+P_1^1\big/\left(8.5-0.25P_1^1\right)\right] + \log_2\left[1+\left(10-P_1^1\right)\big/\left(1.5+0.25P_1^1\right)\right]$ bits. Because $\dfrac{\partial R_1}{\partial P_1^1} < 0$, $R_1$ is maximized when $P_1^1 = 0$. The achievable rates attained at the Stackelberg equilibrium is $R_1^{SE} \approx 2.939$ bits and $R_2^{SE} \approx 3.474$ bits. The unique Nash equilibrium is reached by $\boldsymbol{P}_1^{NE} = \{2, 8\}$ and $\boldsymbol{P}_2^{NE} = \{8, 2\}$ and its achievable rates are $R_1^{NE} = R_2^{NE} \approx 2.645$ bits.

***Remark 2*:** For some channel realizations, the Nash strategy solves the problem (10). If $\alpha_i^f = 0$ for $\forall i, f$, the upper-level and lower-level problems in bi-level program (10) are reduced to two uncoupled problems and the single user water-filling solution can achieve the upper bound in (8). In addition, we give a non-trivial example in which $\alpha_i^f \neq 0$ for $\forall i, f$ and the Nash strategy still solves the problem in (10).

***Example 2*:** Set the parameters $N_1^1, N_2^2$ in Example 1 to be 6, and keep the remaining ones unchanged. We have $R_1 = \log_2\left[1 + P_1^1 \big/ \left(11 - 0.25P_1^1\right)\right] + \log_2\left[1 + \left(10 - P_1^1\right) \big/ \left(1 + 0.25P_1^1\right)\right]$ bits. In this channel realization, the Nash equilibrium coincides with the Stackelberg equilibrium. Both equilibria are reached at $\boldsymbol{P}_1 = \{0, 10\}$ and $\boldsymbol{P}_2 = \{10, 0\}$ and the resulting rates are $R_1 = R_2 \approx 3.460$ bits.

***Remark 3*:** As opposed to the narrow-band case [15], we would like to highlight that the degrees of freedom in allocating the power across multiple bands is essential for the foresighted user to improve its performance. Consider the single-band case in which $N = 1$. Note that user $i$ 's achievable rate $R_i$ is monotonically increasing in its transmitted power $P_i$. If users selfishly maximize their achievable rates, all of them will transmit at their maximum power in the single band, which results in the unique Nash equilibrium. It is easy to check that it is also the unique Stackelberg equilibrium and it is also Pareto efficient.

Although these examples provide us some intuition about the relationship between NE and SE, we are still interested in computing the Stackelberg equilibrium in general scenarios. The following subsection will reformulate the bi-level program into a single-level problem, which helps us to understand the computational complexity of the Stackelberg equilibrium in the multi-user power control games.

## *B. An Exact Single-level Reformulation*

Bi-level programming problems belong to the mathematical programs having optimization problems as constraints. It is well-known they are intrinsically difficult to solve [14]. To understand the computational complexity, we first transform the original bi-level program into a single-level reformulation with the form of

$$\begin{aligned} &\max_{\boldsymbol{P}_1} \sum_{f=1}^{N} \log_2\left(1 + \frac{P_1^f}{N_1^f + \alpha_2^f g_2^f\left(\boldsymbol{P}_1, \boldsymbol{N}_2, \boldsymbol{\alpha}_1, \mathrm{P}_{\mathbf{2}}^{\mathbf{max}}\right)}\right) \\ &s.t.\ \sum\nolimits_{f=1}^{N} P_1^f \leq \mathrm{P}_{\mathbf{1}}^{\mathbf{max}},\ P_1^f \geq 0. \end{aligned} \tag{11}$$

in which $g_2^f\left(\boldsymbol{P}_1,\boldsymbol{N}_2,\boldsymbol{\alpha}_1,\mathrm{P}_{\mathbf{2}}^{\mathbf{max}}\right)$ is a function that determines user 2's allocated power in the *f*th channel, $\boldsymbol{N}_2=\left\{N_2^1,N_2^2,\cdots,N_2^N\right\}$, and $\boldsymbol{\alpha}_1=\left\{\alpha_1^1,\alpha_1^2,\cdots,\alpha_1^N\right\}$.

Note that the lower-level problem in (10) is a standard convex programming problem. Its optimum is given by $P_2^f=g_2^f\left(\boldsymbol{P}_1,\boldsymbol{N}_2,\boldsymbol{\alpha}_1,\mathrm{P}_{\mathbf{2}}^{\mathbf{max}}\right)=\left(K_2-N_2^f-\alpha_1^f P_1^f\right)^+$, where $K_2$ is a constant that satisfies $\sum_{f=1}^N P_2^f=\mathrm{P}_{\mathbf{2}}^{\mathbf{max}}$. In practice, $K_2$ is usually obtained using numerical (e.g. bisection) methods. In fact, an explicit expression of $g_2^f\left(\boldsymbol{P}_1,\boldsymbol{N}_2,\boldsymbol{\alpha}_1,\mathrm{P}_{\mathbf{2}}^{\mathbf{max}}\right)$ is needed to analytically handle single-level formulation. Towards this end, we first define a permutation $\pi:\{1,2,\cdots,N\}\rightarrow\{1,2,\cdots,N\}$, which ranks all the channels based on their noise plus interference PSDs and satisfies

$$\pi(f_1)<\pi(f_2),\ \ if\ N_2^{f_1}+\alpha_1^{f_1}P_1^{f_1}<N_2^{f_2}+\alpha_1^{f_2}P_1^{f_2}. \tag{12}$$

Then, we can extend the results in [16], and have the following closed-form expression:

$$g_2^f\left(\boldsymbol{P}_1,\boldsymbol{N}_2,\boldsymbol{\alpha}_1,\mathrm{P}_{\mathbf{2}}^{\mathbf{max}}\right)=\begin{cases}\dfrac{1}{k}\left(\mathrm{P}_{\mathbf{2}}^{\mathbf{max}}+\sum\limits_{m=1}^{k}\left(N_2^{\pi^{-1}(m)}+\alpha_1^{\pi^{-1}(m)}P_1^{\pi^{-1}(m)}\right)\right)-N_2^f-\alpha_1^f P_1^f, & \pi(f)\le k,\\ 0, & \pi(f)>k,\end{cases} \tag{13}$$

where $k$ can be found according to the condition:

$$\begin{aligned}&k\left(N_2^{\pi^{-1}(k)}+\alpha_1^{\pi^{-1}(k)}P_1^{\pi^{-1}(k)}\right)-\sum_{m=1}^{k}\left(N_2^{\pi^{-1}(m)}+\alpha_1^{\pi^{-1}(m)}P_1^{\pi^{-1}(m)}\right)<\mathrm{P}_{\mathbf{2}}^{\mathbf{max}}\le\\&(k+1)\left(N_2^{\pi^{-1}(k+1)}+\alpha_1^{\pi^{-1}(k+1)}P_1^{\pi^{-1}(k+1)}\right)-\sum_{m=1}^{k+1}\left(N_2^{\pi^{-1}(m)}+\alpha_1^{\pi^{-1}(m)}P_1^{\pi^{-1}(m)}\right).\end{aligned} \tag{14}$$

We can see that function $g_2^f\left(\boldsymbol{P}_1,\boldsymbol{N}_2,\boldsymbol{\alpha}_1,\mathrm{P}_{\mathbf{2}}^{\mathbf{max}}\right)$ ranks all the frequency channels based on the channel conditions and gradually increases the water-level until the maximal power constraint is satisfied.

Even though we have the closed-form expression of $g_2^f\left(\boldsymbol{P}_1,\boldsymbol{N}_2,\boldsymbol{\alpha}_1,\mathrm{P}_{\mathbf{2}}^{\mathbf{max}}\right)$, the single-level problem (11) is still intractable due to its non-convexity. Generally speaking, the global optimum can only be found via an exhaustive search. If we define the granularity in the foresighted user's transmit power as $\Delta_P$, then the value of $P_1^f$ can be limited to the set $\left\{0,\Delta_P,\cdots,\mathbf{P}_{\mathbf{1}}^{\mathbf{max}}\right\}$. By searching all the possible combinations, the optimum can be found. Hence, such an exhaustive search in $\left(P_1^1,\cdots,P_1^N\right)$ has a overall complexity of $\mathcal{O}((\mathbf{P}_{\mathbf{1}}^{\mathbf{max}}/\Delta_P)^N)$.

Recently, Lagrangian duality theory has been successfully used to solve non-convex weighted sum-rate maximization in interference channel with moderate computational complexity [4]-[6]. We notice that the problem in (11) are similar with the problems investigated in these works in that the optimization variables $\boldsymbol{P}_1$ also appear in the denominators of the objective function. The following sections will revisit these dual approaches and show that these methods cannot reduce the computational complexity of problem (11), thereby demonstrating the challenges involved in optimally computing the Stackelberg equilibrium.

### C. Lagrangian Dual Approach for Non-convex Problems

We continue studying the simple two-user scenario to introduce the dual method. In a two-user frequency-selective interference channel, the weighted sum-rate maximization investigated in [4]-[6] is given by

$$\begin{aligned}&\max_{\boldsymbol{P}_1,\boldsymbol{P}_2} \omega\sum_{f=1}^{N}\log_2\left(1+\frac{P_1^f}{N_1^f+\alpha_2^f P_2^f}\right)+(1-\omega)\sum_{f=1}^{N}\log_2\left(1+\frac{P_2^f}{N_2^f+\alpha_1^f P_1^f}\right)\\ &s.t.\ \sum\nolimits_{f=1}^{N} P_1^f\le \mathrm{P}_{\mathbf{1}}^{\mathbf{max}},\ P_1^f\ge 0,\ \sum\nolimits_{f=1}^{N} P_2^f\le \mathrm{P}_{\mathbf{2}}^{\mathbf{max}},\ P_2^f\ge 0,\end{aligned} \tag{15}$$

in which $\omega\in[0,1]$ is a fixed weight. The dual method forms the following Lagrangian

$$L(\boldsymbol{P}_1,\boldsymbol{P}_2,\lambda_1,\lambda_2)=\sum_{f=1}^{N}\left\{\omega\log_2\left(1+\frac{P_1^f}{N_1^f+\alpha_2^f P_2^f}\right)+(1-\omega)\log_2\left(1+\frac{P_2^f}{N_2^f+\alpha_1^f P_1^f}\right)-\lambda_1 P_1^f-\lambda_2 P_2^f\right\}, \tag{16}$$

where $\lambda_1,\lambda_2\ge 0$ are Lagrangian dual variables. The Lagrangian dual function is defined as

$$D(\lambda_1,\lambda_2)=\max_{\boldsymbol{P}_1,\boldsymbol{P}_2\succeq\mathbf{0}} L(\boldsymbol{P}_1,\boldsymbol{P}_2,\lambda_1,\lambda_2). \tag{17}$$

Denote the objective function of problem (15) as $f(\boldsymbol{P}_1,\boldsymbol{P}_2)$ and the overall complexity of exhaustive search is $\mathcal{O}((\prod_i(\mathbf{P_i^{max}}/\Delta_P))^N)$. From optimization theory [17], we know that, for arbitrary feasible $\boldsymbol{P}_1,\boldsymbol{P}_2$, we have $f(\boldsymbol{P}_1,\boldsymbol{P}_2)\le D(\lambda_1,\lambda_2)$. This leads to $\min_{\lambda_1,\lambda_2} D(\lambda_1,\lambda_2)\ge \max_{\boldsymbol{P}_1,\boldsymbol{P}_2} f(\boldsymbol{P}_1,\boldsymbol{P}_2)$, and $\min_{\lambda_1,\lambda_2} D(\lambda_1,\lambda_2)$ provides an upper bound of the optimal value of the problem in (15). Generally speaking, if $f(\boldsymbol{P}_1,\boldsymbol{P}_2)$ is non-convex, the duality gap $\min_{\lambda_1,\lambda_2} D(\lambda_1,\lambda_2)-\max_{\boldsymbol{P}_1,\boldsymbol{P}_2} f(\boldsymbol{P}_1,\boldsymbol{P}_2)$ is not zero.

Fig. 3 summarizes the three key steps of a dual method, the OSB algorithm [4][5], that can efficiently find the global optimum of the problem in (15). First, for fixed $\lambda_1,\lambda_2$, the maximization of $L(\boldsymbol{P}_1,\boldsymbol{P}_2,\lambda_1,\lambda_2)$ over $\boldsymbol{P}_1,\boldsymbol{P}_2$ in (17) is decomposed into $N$ uncoupled sub-problems, and each of them corresponds to a per-bin optimization. Therefore, the overall complexity of maximizing $L(\boldsymbol{P}_1, \boldsymbol{P}_2, \lambda_1, \lambda_2)$ is only $\mathcal{O}(N\prod_i (\mathbf{P_i^{max}}/\Delta_P))$. Second, it is shown that, for fixed $k$, the sum power of user $k$'s optimal power allocation in a multicarrier system is a monotonic function of $\lambda_k$ (Lemma 1, in [4]). This property guarantees that the bi-section dual update over $\lambda_1, \lambda_2$ will converge to the dual optimum. Third, it is also proven that, if the number of frequency bins $N$ is large enough and $H_{ij}^f$ and $\sigma_k^f$ are smooth in the spectral domain, the optimization problem (15) satisfies the so-called "time-sharing property" (Theorem 1 and 2, in [5]), and the duality gap of this non-convex problem is zero. Combining the three properties together, the dual approach can find the global optimum with the computational complexity of $\mathcal{O}(T_1 N\prod_i (\mathbf{P_i^{max}}/\Delta_P))$, where $T_1$ is the number of iterations needed for dual-update. We can see that the complexity of the dual approach is greatly reduced compared with that of the exhaustive search in the primal domain. In addition, it is found in [5] that, if $D(\lambda_1, \lambda_2)$ is approximated using a local maximum of $L(\boldsymbol{P}_1, \boldsymbol{P}_2, \lambda_1, \lambda_2)$, the ISB algorithm can achieve near-optimal performance with the computational complexity of $\mathcal{O}(T_1 T_2 N\sum_i (\mathbf{P_i^{max}}/\Delta_P))$, where $T_2$ is the number of iterations required for evaluating the local maximum.

## *D. The Lagrangian Dual Approach for Computing Stackelberg Equilibrium*

Now we apply the dual approach for our problem in (11) to understand why the Stackelberg equilibrium in our considered problem is intrinsically difficult to compute. Fig. 4 summarizes the key properties of the dual approach that will be addressed in the following parts. Denote the objective function of problem (11) as $f'(\boldsymbol{P}_1)$. Consider its dual objective function $D'(\mu)$ for a fixed Lagrangian dual variable $\mu$:

$$D'(\mu) = \max_{\boldsymbol{P}_1 \succeq \mathbf{0}} L'(\boldsymbol{P}_1, \mu) \tag{18}$$

in which $L'(\boldsymbol{P}_1, \mu) = \sum_{f=1}^{N} \log_2\left(1 + \frac{P_1^f}{N_1^f + \alpha_2^f g_2^f(\boldsymbol{P}_1, \boldsymbol{N}_2, \boldsymbol{\alpha}_1, \mathrm{P}_\mathbf{2}^\mathbf{max})}\right) + \mu\left(\mathrm{P}_\mathbf{1}^\mathbf{max} - \sum_{f=1}^{N} P_1^f\right)$. For a given $\mu$, denote the optimal power allocation that maximizes (18) as $\boldsymbol{P}_1(\mu) = \arg\max_{\boldsymbol{P}_1 \succeq \mathbf{0}} L'(\boldsymbol{P}_1, \mu)$ and $\boldsymbol{P}_1^f(\mu) = [\boldsymbol{P}_1(\mu)]_f$, The following lemma holds for $\boldsymbol{P}_1(\mu)$:

***Lemma 1***: $\sum_{f=1}^{N} \boldsymbol{P}_1^f(\mu)$ is monotonic decreasing in $\mu$. In addition, we have $\lim_{\mu\to\infty}\sum_{f=1}^{N} \boldsymbol{P}_1^f(\mu)=0$ and $\sum_{f=1}^{N} \boldsymbol{P}_1^f(0)=+\infty$.

***Proof*** : It is easy to see that $\sum_{f=1}^{N} \boldsymbol{P}_1^f(0)=+\infty$. The rest of the proof is the same as in Lemma 1 in [4].■

Fig. 5 gives a graphical illustration of the above Lemma. Consider a sequence of optimization problems similar with (11). These problems are parameterized by the constraint imposed over user 1's maximal sum power. The solid curve in Fig. 5 is a plot of the optimal value $\left(\sum_{f=1}^{N} P_1^{*f}, f'\left(\boldsymbol{P}_1^*\right)\right)$ as this constraint varies. The curve is plotted with $\sum_{f=1}^{N} P_1^{*f}$ on the *x*-axis. The *y*-axis is located at the point where $\sum_{f=1}^{N} P_1^{*f} = \mathrm{P}_1^{\mathbf{max}}$. The intersection of the curve with the *y*-axis is the optimum of (11), i.e. $\max_{\boldsymbol{P}_1} f'\left(\boldsymbol{P}_1\right)$. For a fixed $\mu$, by drawing a tangent line to the $\left(\sum_{f=1}^{N} P_1^{*f}, f'\left(\boldsymbol{P}_1^*\right)\right)$ curve and measuring the intersection of this tangent line with the *y*-axis, the value of $D'(\mu)$ can be graphically obtained. According to Lemma 1, as $\mu$ increases, the *x*-axis value of the tangent point monotonically increases. We denote $\mu^* = \arg\min_{\mu} D'(\mu)$. Recall that Lemma 1 does not claim the continuity of $\sum_{f=1}^{N} \boldsymbol{P}_1^f(\mu)$ in $\mu$. It is because the allocated powers in different frequency bins are coupled due to function $g_2^f\left(\boldsymbol{P}_1, \boldsymbol{N}_2, \boldsymbol{\alpha}_1, \mathrm{P}_{\mathbf{2}}^{\mathbf{max}}\right)$ and the time-sharing property in [5] is not guaranteed for problem (11). The discontinuity may lead to nonzero duality gap, i.e. at least two tangent points exist on the tangent line in Fig. 5 and they correspond to different power constraints $\mathrm{P}_1^{\mathbf{x}}$ and $\mathrm{P}_1^{\mathbf{y}}$. If the duality gap is positive, the following theorem indicates that $D'\left(\mu^*\right)$ provides a tighter upper bound of the achievable rate than $R_1^{\max}$ in (8).

***Theorem 2***: If the duality gap is nonzero, i.e. $D'\left(\mu^*\right) > \max_{\boldsymbol{P}_1} f'\left(\boldsymbol{P}_1\right)$, the dual optimum provides a tighter upper bound of user 1's maximal achievable rate than the bound in (8), i.e. $D'\left(\mu^*\right) < R_1^{\max}$.

***Proof*** : As shown in Fig. 5, the non-zero duality gap implies that there exist at least two possible values for $\sum_{f=1}^{N} \boldsymbol{P}_1^f\left(\mu^*\right)$, which are denoted as $\mathrm{P}_1^{\mathbf{x}}$ and $\mathrm{P}_1^{\mathbf{y}}$ and they satisfy $\mathrm{P}_1^{\mathbf{x}} < \mathrm{P}_1^{\mathbf{max}} < \mathrm{P}_1^{\mathbf{y}}$. Denote the optimal power allocation of having power constraints $\mathrm{P}_1^{\mathbf{x}}$ and $\mathrm{P}_1^{\mathbf{y}}$ as $\boldsymbol{P}_1^-$ and $\boldsymbol{P}_1^+$ respectively. We have that

$$D'\left(\mu^*\right)=f'\left(\boldsymbol{P}_1^{-}\right)+\mu^*\left(\mathrm{P}_{\mathbf{1}}^{\mathbf{max}}-\sum_{f=1}^{N}P_1^{-f}\right)=f'\left(\boldsymbol{P}_1^{+}\right)+\mu^*\left(\mathrm{P}_{\mathbf{1}}^{\mathbf{max}}-\sum_{f=1}^{N}P_1^{+f}\right). \tag{19}$$

Moreover, since $\mathrm{P}_{\mathbf{1}}^{\mathbf{x}}<\mathrm{P}_{\mathbf{1}}^{\mathbf{max}}<\mathrm{P}_{\mathbf{1}}^{\mathbf{y}}$, there exists $0<\upsilon<1$ such that $\mathrm{P}_{\mathbf{1}}^{\mathbf{max}}=\upsilon\mathrm{P}_{\mathbf{1}}^{\mathbf{x}}+(1-\upsilon)\mathrm{P}_{\mathbf{1}}^{\mathbf{y}}$. Immediately, we get $D'\left(\mu^*\right)=\upsilon f'\left(\boldsymbol{P}_1^{-}\right)+(1-\upsilon)f'\left(\boldsymbol{P}_1^{+}\right)$. It corresponds to the time-sharing scenario, in which the power allocation $\boldsymbol{P}_1^{-}$ is adopted for time-fraction $\upsilon$ and $\boldsymbol{P}_1^{+}$ for time-fraction $1-\upsilon$. Consider the problem of allocating user 1's power subject to the maximal power constraint $\mathrm{P}_{\mathbf{1}}^{\mathbf{max}}$ in the interference-free environment. We know that the optimal solution is the single-user water-filling. Noting that $\mathrm{P}_{\mathbf{1}}^{\mathbf{max}}=\upsilon\mathrm{P}_{\mathbf{1}}^{\mathbf{x}}+(1-\upsilon)\mathrm{P}_{\mathbf{1}}^{\mathbf{y}}$ and $\mathrm{P}_{\mathbf{1}}^{\mathbf{x}}\neq\mathrm{P}_{\mathbf{1}}^{\mathbf{y}}$, the aforementioned time-sharing strategy is sub-optimal for this problem. Therefore, we have

$$\begin{aligned}D'\left(\mu^*\right)&=\upsilon f'\left(\boldsymbol{P}_1^{-}\right)+(1-\upsilon)f'\left(\boldsymbol{P}_1^{+}\right)\le\upsilon\sum_{f=1}^{N}\log_2\left(1+P_1^{-f}\left|H_{11}^f\right|^2\Big/\sigma_1^f\right)\\&+(1-\upsilon)\sum_{f=1}^{N}\log_2\left(1+P_1^{+f}\left|H_{11}^f\right|^2\Big/\sigma_1^f\right)<\sum_{f=1}^{N}\log_2\left(1+P_1^{f*}\left|H_{11}^f\right|^2\Big/\sigma_1^f\right)=R_1^{\max},\end{aligned} \tag{20}$$

and this concludes the proof. ■

By Theorem 2, evaluating the dual function leads to a tighter upper bound of Stackelberg equilibrium than $R_1^{\max}$. However, it is unfortunate that the computational complexity of optimally maximizing $L'\left(\boldsymbol{P}_1,\mu\right)$ is still $\mathcal{O}((\mathbf{P}_{\mathbf{1}}^{\mathbf{max}}/\Delta_P)^N)$. This is because term $g_2^f\left(\boldsymbol{P}_1,\boldsymbol{N}_2,\boldsymbol{\alpha}_1,\mathrm{P}_{\mathbf{2}}^{\mathbf{max}}\right)$ in the denominator term of (11) is also a function of the allocated power $P_1^{f'}\left(f'\neq f\right)$, which makes it impossible to decouple the maximization in (18) into $N$ independent sub-problems. To conclude, the complexity of optimal solution in the dual domain is the same as the primal approach, which again highlights the fact that the Stackelberg equilibrium is difficult to compute.

# IV. Low-complexity Algorithm, Simulations, and Extensions

In this section, we propose a low-complexity dual algorithm to search the Stackelberg equilibrium and examine its achievable performance via extensive numerical simulations. We also discuss how the strategic users can obtain the required information and the extensions to general multi-user scenarios.

## A. A Low-Complexity Dual Approach

As we have shown, the dual approach cannot reduce the complexity of the global optimum of problem (11).

However, inspired by the ISB algorithm [5], we develop an efficient dual approach, which is listed as Algorithm 1. The basic idea of the algorithm is to approximately evaluate $D'(\mu)$ by locally optimizing $L'(\boldsymbol{P}_1, \mu)$. For fixed $\mu$, the algorithm finds the optimal $P_1^f$ while keeping $P_1^1, \cdots, P_1^{f-1}, P_1^{f+1}, \cdots, P_1^N$ fixed, and changes the index $f$ until it converges to a local maximum for $L'(\boldsymbol{P}_1, \mu)$. Then the algorithm updates $\mu$ using bi-section search and repeats the procedure above until the convergence is achieved.

As discussed in [5], the local optimum depends on the initial starting point and the ordering of iterations. Moreover, the proof of convergence of the whole algorithm becomes an issue. Algorithm 1 sets the Nash equilibrium as the initial starting point. In most of the experimental setting we have tested, Algorithm 1 has been observed to converge to a feasible solution within 10-15 iterations. The computational complexity of this iterative algorithm is only $\mathcal{O}(T_1 T_2 N \, \mathbf{P}_1^{\max} / \Delta_P)$ and it reduces the complexity of the optimal exhaustive search by a factor of $\mathcal{O}((\mathbf{P}_1^{\max} / \Delta_P)^{N-1} / (T_2 N))$, which is considerably large for small $\Delta_P$ and large $N$. Table I summarizes the computational complexity comparison for user 1 if it adopts different algorithms, in which $T_3$ is the number of iterations required in the iterative water-filling algorithm.

## B. *Illustrative Results*

In this sub-section, we evaluate the performance of Algorithm 1 by comparing with the IW algorithm. We simulate a wireless system with 20 sub-carriers over the 6.25-MHz band. We assume that $\mathrm{P}_1^{\max} = \mathrm{P}_2^{\max} = 200$ and $\sigma_1^f = \sigma_2^f = 0.01$. To evaluate the performance, we tested $10^5$ sets of frequency-selective fading channels where a unique Nash equilibrium exists, which are simulated using a four-ray Rayleigh model with the exponential power profile and 160 ns delay between two adjacent rays [18]. The simulated power of each ray decreases exponentially according to its delay. The total power of all rays of $H_{11}^f$ and $H_{22}^f$ is normalized as one, and that of $H_{12}^f$ and $H_{21}^f$ is normalized as $0.5$.

Fig. 6 and 7 show the power allocations for both users using different algorithms. In the IW algorithm, each user water-fills the whole frequency band by regarding its competitor's transmit PSD as background noise until the Nash equilibrium is achieved. In contrast, user 1 does not water-fill if it adopts Algorithm 1. For example, in Fig. 6, user 1 allocates a large amount of power in frequency bin 3 even though it can gain an immediate increase in $R_1$ by re-allocating some of its power in the frequency bins 5 and 6 where the noise plus interference PSD is below its water-levels in the frequency bins 7-12.

Denote user $i$'s achieved rate by deploying Algorithm 1 as $R_i'$. Fig. 8 shows the simulated cumulative distribution functions (cdf) of $R_i' / R_i^{NE}$. From the curve, Algorithm 1 achieves a higher rate for the foresighted user in all the simulated realizations. The average rate improvement that Algorithm 1 provides over the IW algorithm is 38%. In addition, it is surprising to find that, in 95% of the simulation settings, Algorithm 1 also results in a higher rate $R_2'$ than $R_2^{NE}$ for the myopic user, and the average rate improvement is 45%. This is because user 1's Stackelberg strategy mitigates its interference caused to user 2.

We also simulate the scenarios in which the total power of $H_{12}^f$ and $H_{21}^f$ is normalized as 0.25 and all the other parameters remain the same as above. Fig. 9 shows the simulated cdfs of $R_i' / R_i^{NE}$. The average rate improvement for user 1 is 27% and that of user 2 is 32%. It is intuitive that the average rate improvement is decreasing when the power of $H_{12}^f$ and $H_{21}^f$ decreases, because the interference coupling between users and the foresighted user's ability in shaping the myopic user's response are both reduced.

### C. *Information Acquisition*

Previous sections mentioned that, in order to play the Stackelberg equilibrium, the additional information about the competing user's CSI, maximum power constraint, and power allocation strategy is indispensable. In practice, there are several possible methods to acquire this required information.

First, the myopic user has the incentive to provide the required information, because its performance can be greatly improved if the foresighted player knows the myopic player's private information. In the distributed setting, users can individually decide whether or not to play the Stackelberg strategy based on their computational hardware constraints. The user that wants to behave myopically can reveal its information to the foresighted user. This can be viewed as the user's cooperative behavior to avoid mutual interference.

When no information exchanges among users are possible, the alternative way for users to gather this information is through predictive modeling. If the foresighted user strategically changes its power allocation, it can measure and model the resulting interference PSD, i.e. estimate the functional expression of $g_2^f\left(\boldsymbol{P}_1,\boldsymbol{N}_2,\boldsymbol{\alpha}_1,\mathrm{P}_2^{\max}\right)$, without any information exchange among users. For instance, in [19], we showed that the foresighted user can effectively model its experienced interference as a linear function of its own allocated power, formulate a local approximation of the original bi-level program, and substantially improve both users' achievable rates.

### D. Extensions to Multi-user Games

The two-user formulation can be extended to the general cases in which multiple users can be myopic or foresighted. The analysis in these cases becomes much more involved. We denote the number of foresighted user as $n_f$ and the number of myopic user as $n_m$. We briefly address two remaining cases as follows.

In the first case, $n_f = 1, n_m > 1$. As in (11), we can still have the following single-level formulation:

$$\begin{aligned}&\max_{\boldsymbol{P}_1}\sum_{f=1}^{N}\log_2\left(1+\frac{P_1^f}{N_1^f+\sum_{k=2}^{n_m+1}\alpha_{k1}^f q_k^f\left(\boldsymbol{P}_1,\boldsymbol{N},\boldsymbol{\alpha},\mathrm{P}_2^{\max},\cdots,\mathrm{P}_{n_m+1}^{\max}\right)}\right)\\&s.t.\ \sum_{f=1}^{N}P_1^f\le \mathrm{P}_1^{\max},\ P_1^f\ge 0,\end{aligned}\tag{21}$$

in which $N_i^f=\sigma_i^f\Big/\left|H_{ii}^f\right|^2$, $\boldsymbol{N}=\left\{N_i^f: i=2,\cdots,n_m+1, f=1,\cdots,K\right\}$, $\boldsymbol{\alpha}=\left\{\alpha_{ij}^f: i=1,\cdots,n_m+1,\ j=2,\cdots,n_m+1,\right.$ $\left. f=1,\cdots,K\right\}$, and $q_k^f\left(\boldsymbol{P}_1,\boldsymbol{N},\boldsymbol{\alpha},\mathrm{P}_2^{\max},\cdots,\mathrm{P}_{n_m+1}^{\max}\right)$ is the function determining user *k*'s allocated power in channel *f*. As a general from of the two-user case, problem (21) is also non-convex. It is easy to verify that *Lemma 1* and *Theorem 2* still hold. Although it is difficult to analytically derive $q_k^f\left(\boldsymbol{P}_1,\boldsymbol{N},\boldsymbol{\alpha},\mathrm{P}_2^{\max},\cdots,\mathrm{P}_{n_m+1}^{\max}\right)$, we are still able to numerically evaluate it. Hence, Algorithm 1 can be applied in this case by replacing its lines 7 to 8 with numerically finding local maxima of $\sum_{f=1}^{N}\left(\log_2\left(1+\frac{P_1^f}{N_1^f+\sum_{k=2}^{n_m+1}\alpha_{k1}^f q_k^f\left(\boldsymbol{P}_1,\boldsymbol{N},\boldsymbol{\alpha},\mathrm{P}_2^{\max},\cdots,\mathrm{P}_{n_m+1}^{\max}\right)}\right)-\mu P_1^f\right)$. We simulate the three-user scenarios in which $\sum_{f=1}^{N}\left|H_{ij}^f\right|^2=0.25$ for $i\ne j$, $\mathrm{P}_i^{\max}=200$, and $\sigma_i^f=0.01$, and all the other parameters remain the same as Section IV.B. Fig. 10 shows the simulated cdfs of $R_i'/R_i^{NE}$. The average rate improvement for user 1 is 34% and that of user 2 and 3 is 10.5%. From Fig. 10, we can see that, the Stackelberg strategy also benefits the two myopic users in more than 83% of the channel realizations.

Assume now that we have multiple foresighted users, i.e. $n_f > 1, n_m \geq 1$. In this case, the single objective function in the original upper-level problem disappears and it becomes a multi-objective optimization problem. Using similar arguments in Theorem 1, we can show that the Nash equilibrium still exists in the followers' game. For these foresighted users, a reasonable outcome is to choose an operating point in the set $\mathcal{R}^{n_f} = \left\{\left(R_1,\cdots,R_{n_f}\right): R_i \geq R_i^{NE},\, for\ all\ i=1,\cdots,n_f\right\}$, where $R_i^{NE}$ is user $i$'s achievable rate if all the users are myopic. This point can be determined based on the negotiation among the foresighted users. Cooperative game theory provides many solution concepts, e.g. bargaining, for choosing the operating point [12]. Note that the overall game in this scenario is a mixture of cooperation and competition in that the cooperation exists among the foresighted users while myopic players compete with each other. A possible way of achieving the boundary point on $\mathcal{R}^{n_f}$ is to let some coordinator solve the following weighted sum-rate maximization and determine the transmitted PSDs for different foresighted users:

$$\begin{aligned}&\max_{\boldsymbol{P}_1,\cdots,\boldsymbol{P}_{n_f}} \sum_{f=1}^{N} \omega_i \log_2\left(1+\frac{P_i^f}{N_i^f+\sum_{k=n_f+1}^{n_f+n_m}\alpha_{ki}^f q_k^f\left(\boldsymbol{P}_1,\cdots,\boldsymbol{P}_{n_f},\boldsymbol{N},\boldsymbol{\alpha},\mathrm{P}_{n_f+1}^{\mathbf{max}},\cdots,\mathrm{P}_{n_f+n_m}^{\mathbf{max}}\right)}\right)\\ &s.t.\ \sum_{f=1}^{N} P_i^f \leq \mathrm{P}_{\mathbf{i}}^{\mathbf{max}},\, P_i^f \geq 0,\, R_i \geq R_i^{NE},\ \ i=1,\cdots,n_f,\end{aligned} \tag{22}$$

in which $\omega_i \geq 0$ is user $i$'s weight. Although this problem is generally difficult to solve optimally, some low-complexity methods similar to Algorithm 1 can be adopted to obtain sub-optimal solutions.

## V. CONCLUSIONS

This paper considers the strategic behavior in determining the transmit power PSD for selfish users sharing a frequency-selective interference channel. We adopt the game theoretic concept of Stackelberg equilibrium and model the two-user case as a bi-level programming problem. We show that the Stackelberg equilibrium is intrinsically difficult to compute and propose a low-complexity approach based on Lagrangian dual theory. Numerical results show the strategic user should avoid shortsighted Nash strategy and it can substantially improve both users' performance if it knows the CSI and response strategy of the competing user. Operational methods for acquiring the necessary information and extensions to multi-user scenarios are proposed. Obtaining satisfactory performance with minimal information exchange while multiple foresighted users exist is indentified as a problem for further investigation.

# REFERENCES

[1] W. Yu, G. Ginis, and J. Cioffi, "Distributed multiuser power control for digital subscriber lines," *IEEE J. Sel. Areas Commun.* vol. 20, no. 5, pp.1105-1115, June 2002.

[2] R. Etkin, A. Parekh, and D. Tse, "Spectrum Sharing for Unlicensed Bands," *IEEE J. Sel. Areas Commun.* vol. 25, no. 3, pp. 517-528, April 2007.

[3] J. Huang, R. Berry, and M. Honig, "Distributed Interference Compensation for Wireless Networks," *IEEE J. Sel. Areas Commun.*, vol. 24, no. 5, pp. 1074-1084, May 2006.

[4] R. Cendrillon, W. Yu, M. Moonen, J. Verlinden, and T. Bostoen, "Optimal multiuser spectrum balancing for digital subscriber lines," *IEEE Trans. Commu.*, vol. 54, no. 5, pp. 922-933, May 2006.

[5] W. Yu and R. Lui, "Dual methods for nonconvex spectrum optimization of multicarrier systems", *IEEE Trans. Commu.*, vol.54, p.1310-1322, 2006.

[6] R. Cendrillon, J. Huang, M. Chiang, and M. Moonen, "Autonomous Spectrum Balancing for Digital Subscriber Lines," *IEEE Trans. on Signal Processing*, vol. 55, no. 8, pp. 4241-4257, August 2007.

[7] O. Popescu, D. Popescu, and C. Rose, "Simultaneous Water Filling for Mutually Interfering Systems," *IEEE Trans. Wireless Commu.*, vol. 6, no.3, pp. 1102-1113, Mar. 2007.

[8] S. Haykin,"Cognitive Radio: Brain-empowered wireless communications," *IEEE JSAC*, vol. 23, pp. 201-220, 2005

[9] S. Haykin, "Cognitive Dynamic Systems," *Proc. of IEEE ICASSP 2007*, vol. 4, pp. 1369-1372, April 2007

[10] E. Altman, T. Boulogne, R. El-Azouzi, T Jimenez, and L. Wynter, "A survey on networking games", *Computers and Operations Research*, 2004.

[11] Y. Shoham, R. Powers, and T. Grenager, "If Multi-Agent Learning is the Answer, What is the Question?", *Artificial Intelligence*, vol. 171, no. 7, pp. 365-377, May 2007

[12] D. Fudenberg and J. Tirole, *Game Theory*. Cambridge, MA: MIT Press, 1991.

[13] K. W. Shum, K.-K. Leung, and C. W. Sung, "Convergence of iterative waterfilling algorithm for gaussian interference channels," *IEEE J. Sel. Areas Commun.*, vol. 25, no 6, pp. 1091-1100, Aug. 2007.

[14] B. Colson, P. Marcotte, and G. Savard, "Bilevel programming: A survey," *A quarterly Journal of Operation Research*, vol. 3, no.2, pp. 87-107, 2005.

[15] E. Altman and Z. Altman, "S-modular games and power control in wireless networks," *IEEE Trans. Automatic Control*, vol. 48, no. 5, pp. 839-842, May 2003.

[16] E. Altman, K. Avrachenkov, and A. Garnaev, "Closed form solutions for symmetric water filling games," *Proc. IEEE INFOCOM 2008*, pp. 673-681, April 2008.

[17] S. Boyd and L. Vandenberghe, *Convex Optimization*, Cambridge University Press, 2004.

[18] T. S. Rappaport, *Wireless Communications*. Englewood Cliffs, NJ: Prentice-Hall, 1996.

[19] Y. Su and M. van der Schaar, "Conjectural equilibrium in water-filling games," Technical Report, 2008 (available at http://arxiv.org/abs/0811.0048)

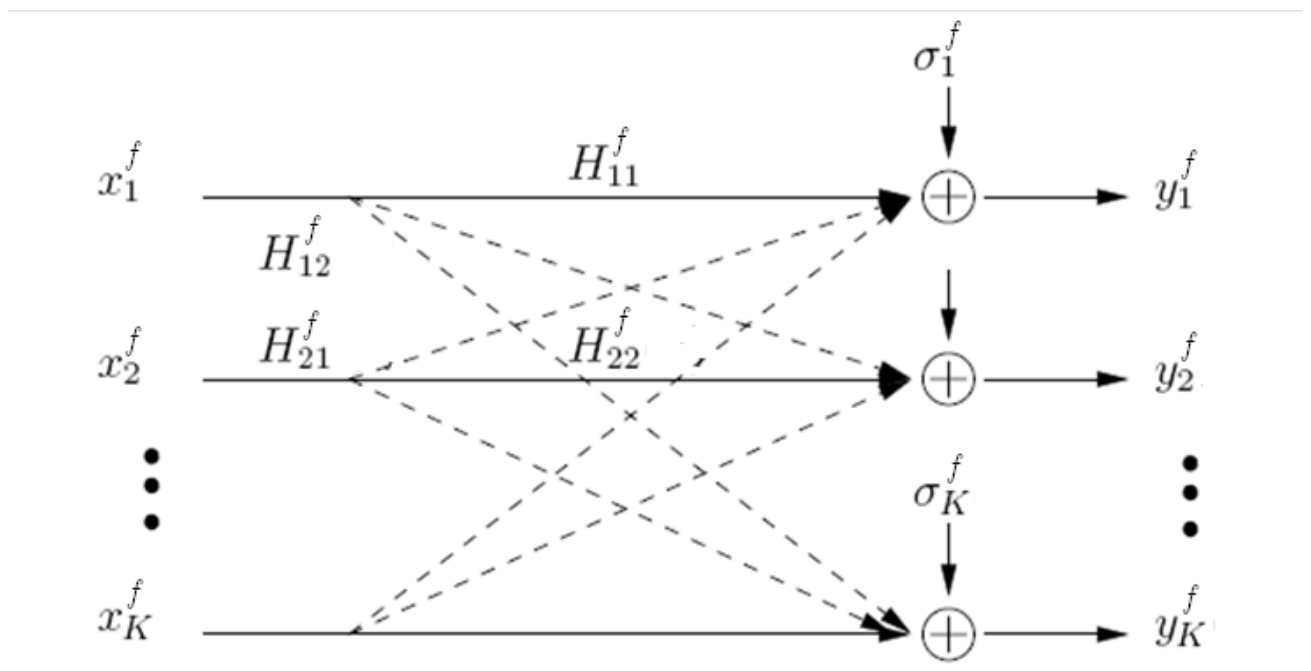

Fig. 1. Gaussian interference channel model.

| | *Left* | *Right* |
|---|---|---|
| *Up* | 1, 0 | 3, 2 |
| *Down* | 2, 1 | 4, 0 |

Fig. 2. Stackelberg game: the row player's payoff is given first in each cell, with the column player's payoff following.

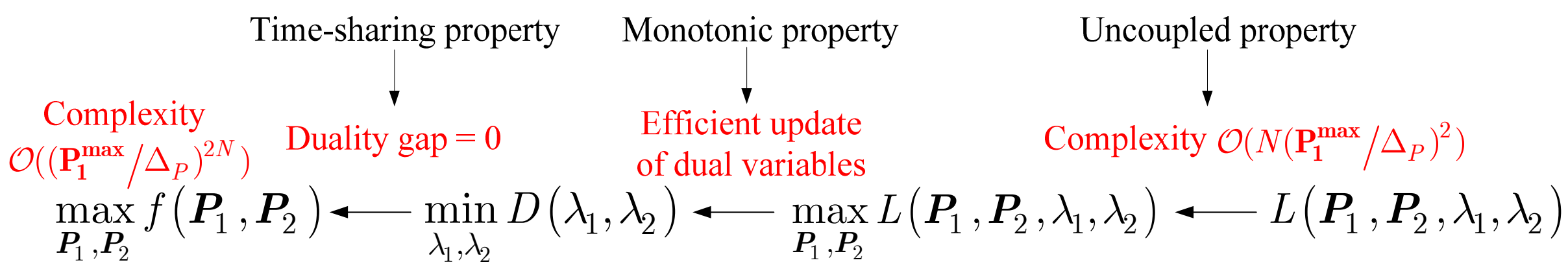

Fig. 3. Key steps of the dual approach of non-convex weighted sum-rate maximization.

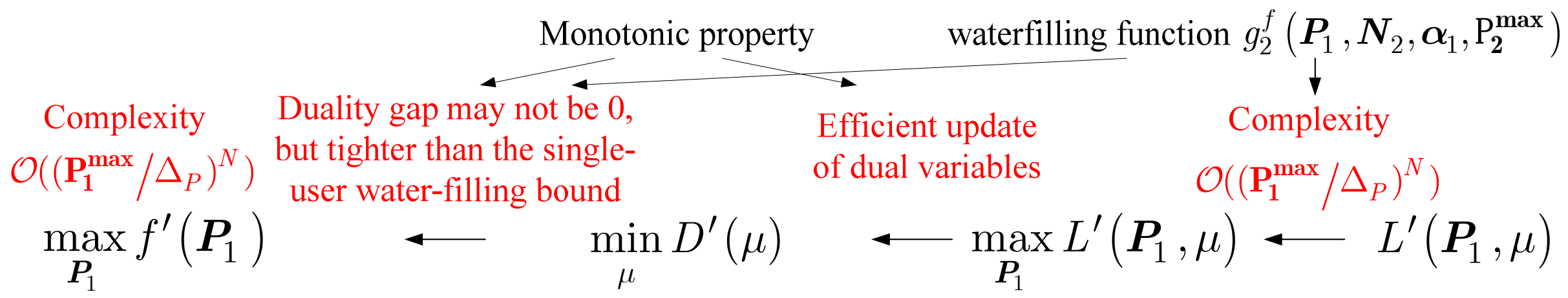

Fig. 4. Complexity and properties of the dual approach of computing the Stackelberg equilibrium.

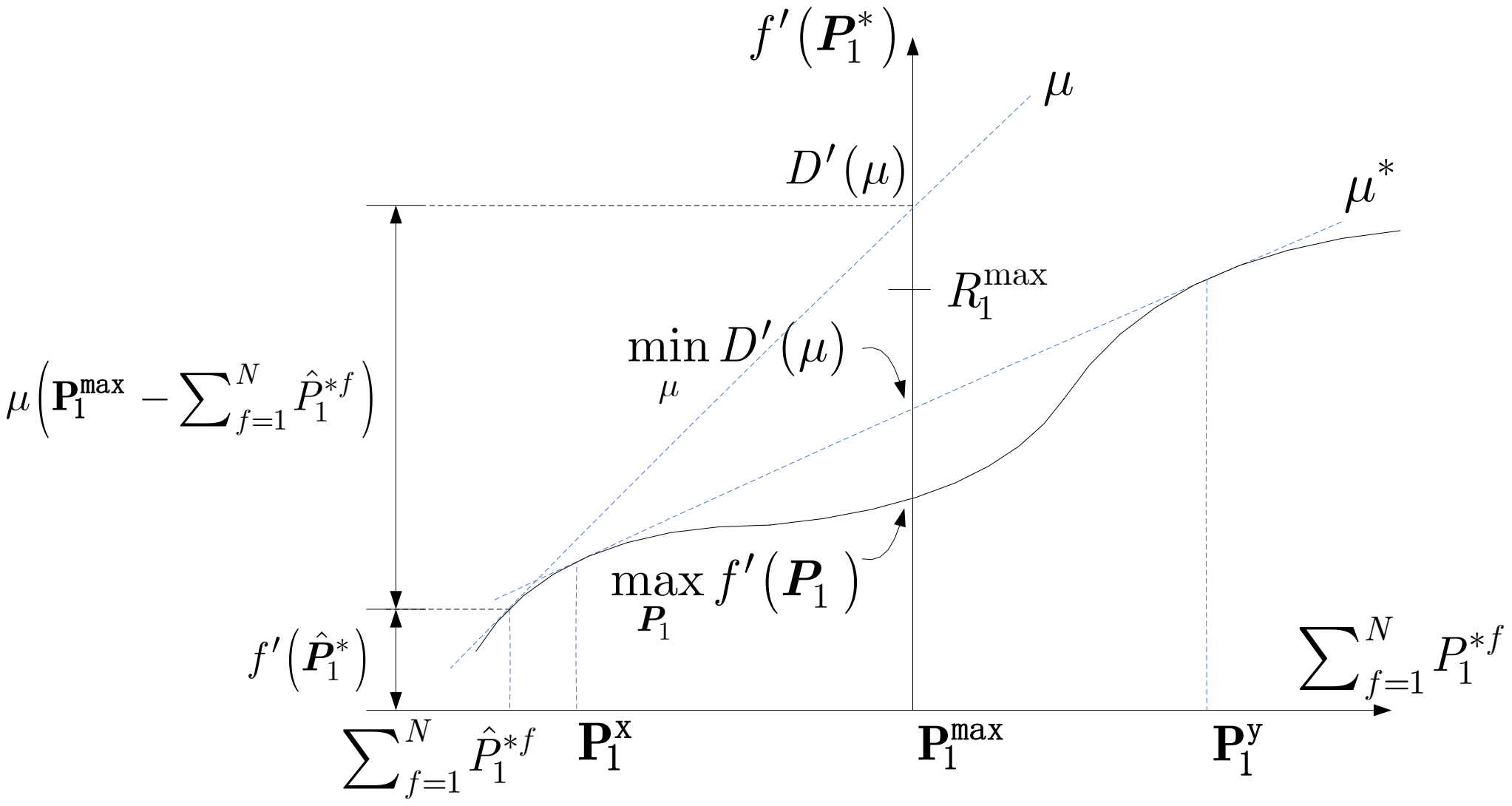

Fig. 5. Duality gap for the problem in (11).

---

**Algorithm 1:** A low-complexity dual approach

---

input: $\mathrm{P}_1^{\mathtt{max}}, \mathrm{P}_2^{\mathtt{max}}, N_1^f, N_2^f, \alpha_1^f, \alpha_2^f$ *for* $\forall f$

initialization : $\boldsymbol{P}_1 = \boldsymbol{P}_1^{NE}, \mu^{\max}, \mu^{\min}$

repeat

$\mu = \left(\mu^{\max} + \mu^{\min}\right)/2.$

repeat

for $f = 1$ to $N$ ,

set $P_1^f = \arg\max_{P_1^f} \sum_{f=1}^{N} \left\{ \ln\left[1 + P_1^f \Big/ \left(N_1^f + \alpha_2^f g_2^f\left(\boldsymbol{P}_1, \boldsymbol{N}_2, \boldsymbol{\alpha}_1, \mathrm{P}_{\mathbf{2}}^{\mathtt{max}}\right)\right)\right] - \mu P_1^f \right\}$

by keeping $P_1^1, \cdots, P_1^{f-1}, P_1^{f+1}, \cdots, P_1^N$ fixed.

end

until $\left(P_1^1, \cdots, P_1^N\right)$ converges

if $\sum_{f=1}^{N} P_1^f > \mathrm{P}_1^{\mathtt{max}}$, $\mu^{\min} = \left(\mu^{\max} + \mu^{\min}\right)/2$ ; else $\mu^{\max} = \left(\mu^{\max} + \mu^{\min}\right)/2$ .

until it converges

---

| Algorithm | Computational complexity |
|---|---|
| Exhaustive search | $\mathcal{O}((\mathbf{P}_1^{\mathbf{max}}/\Delta_P)^N)$ |
| Algorithm 1 | $\mathcal{O}(T_1 T_2 N\, \mathbf{P}_1^{\mathbf{max}}/\Delta_P)$ |
| Iterative water-filling | $\mathcal{O}(T_3 N)$ |

Table I. User 1's computational complexity for different algorithms.

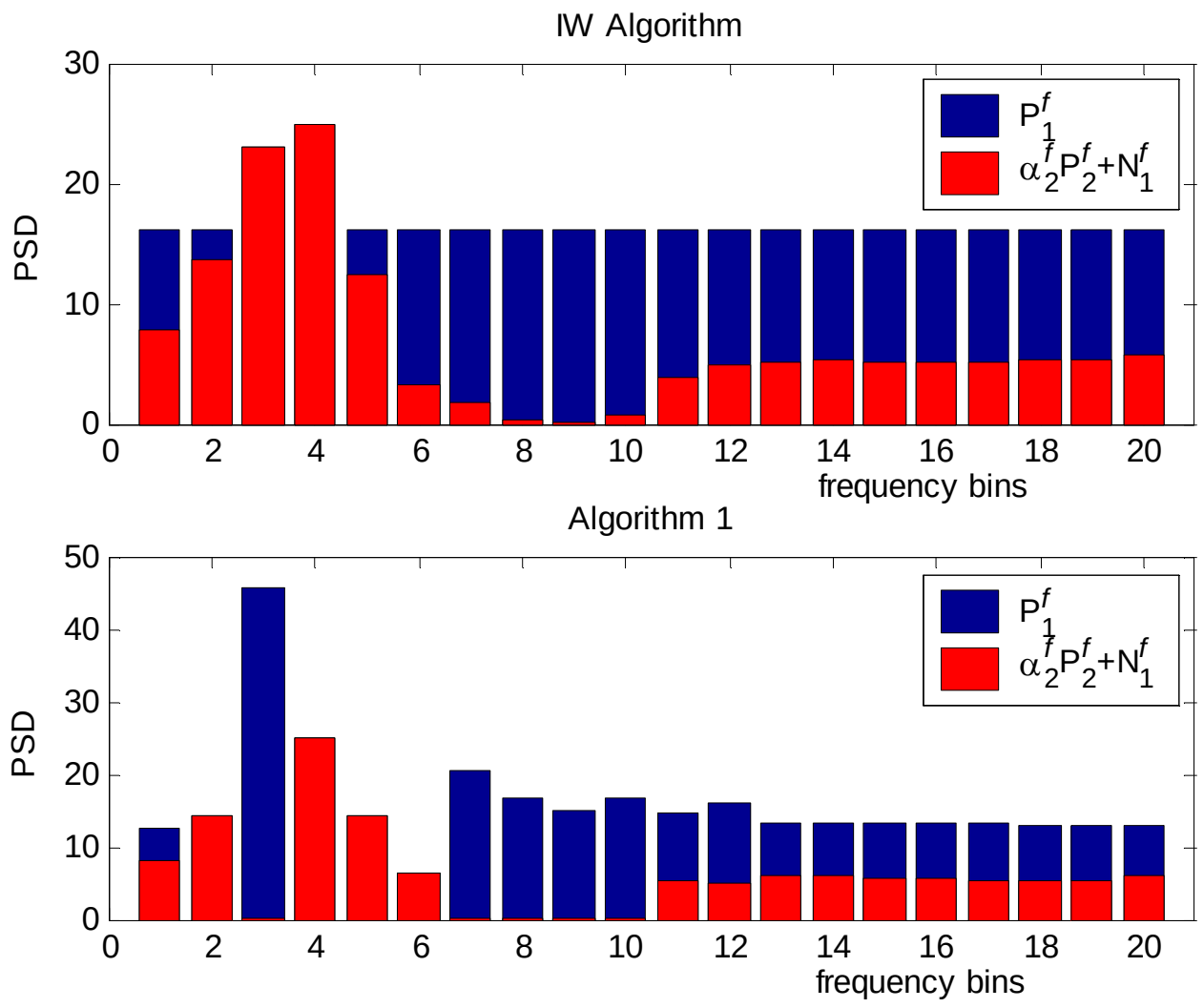

Fig. 6. User 1's power allocation using different algorithms.

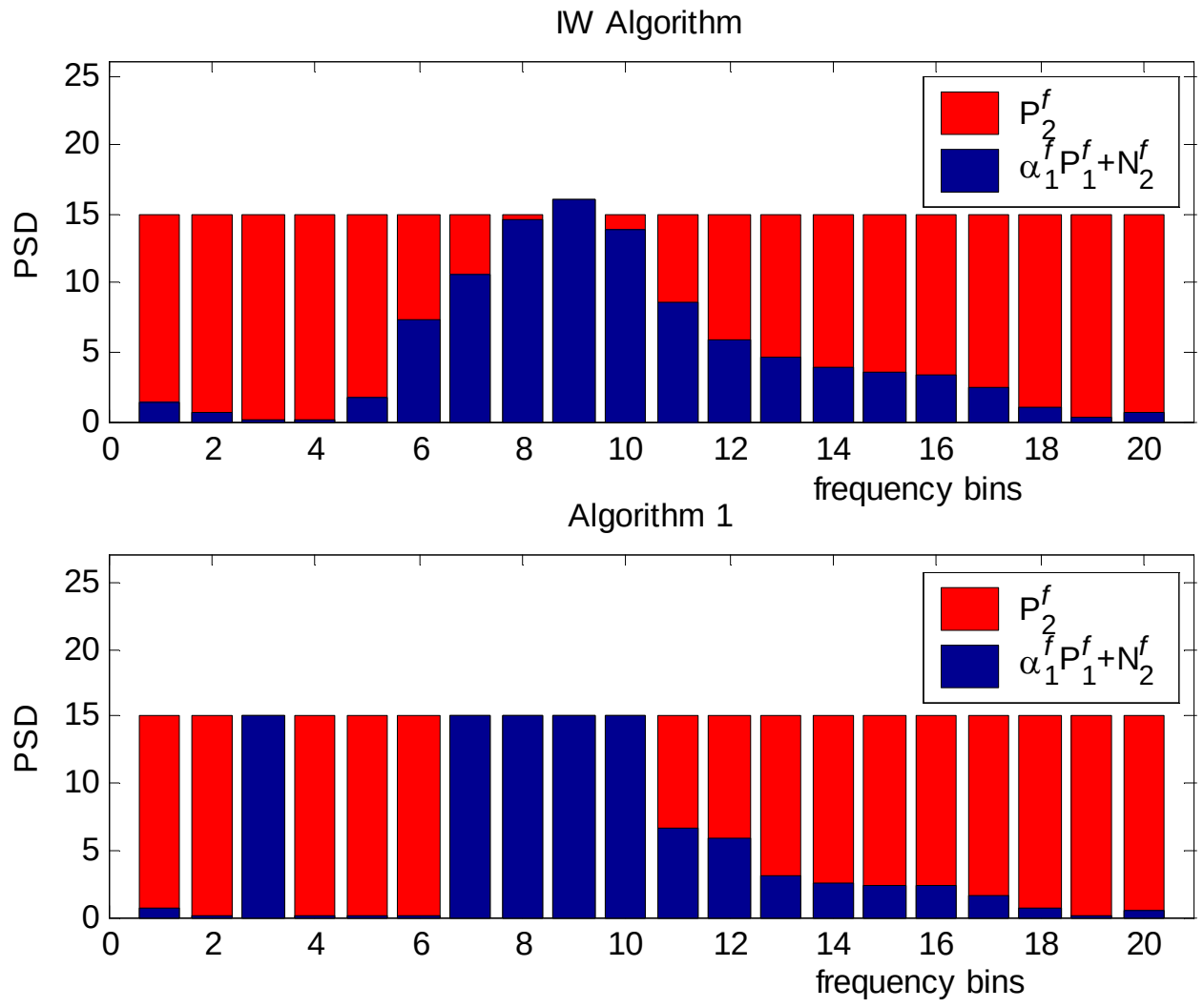

Fig. 7. User 2's power allocation using different algorithms.

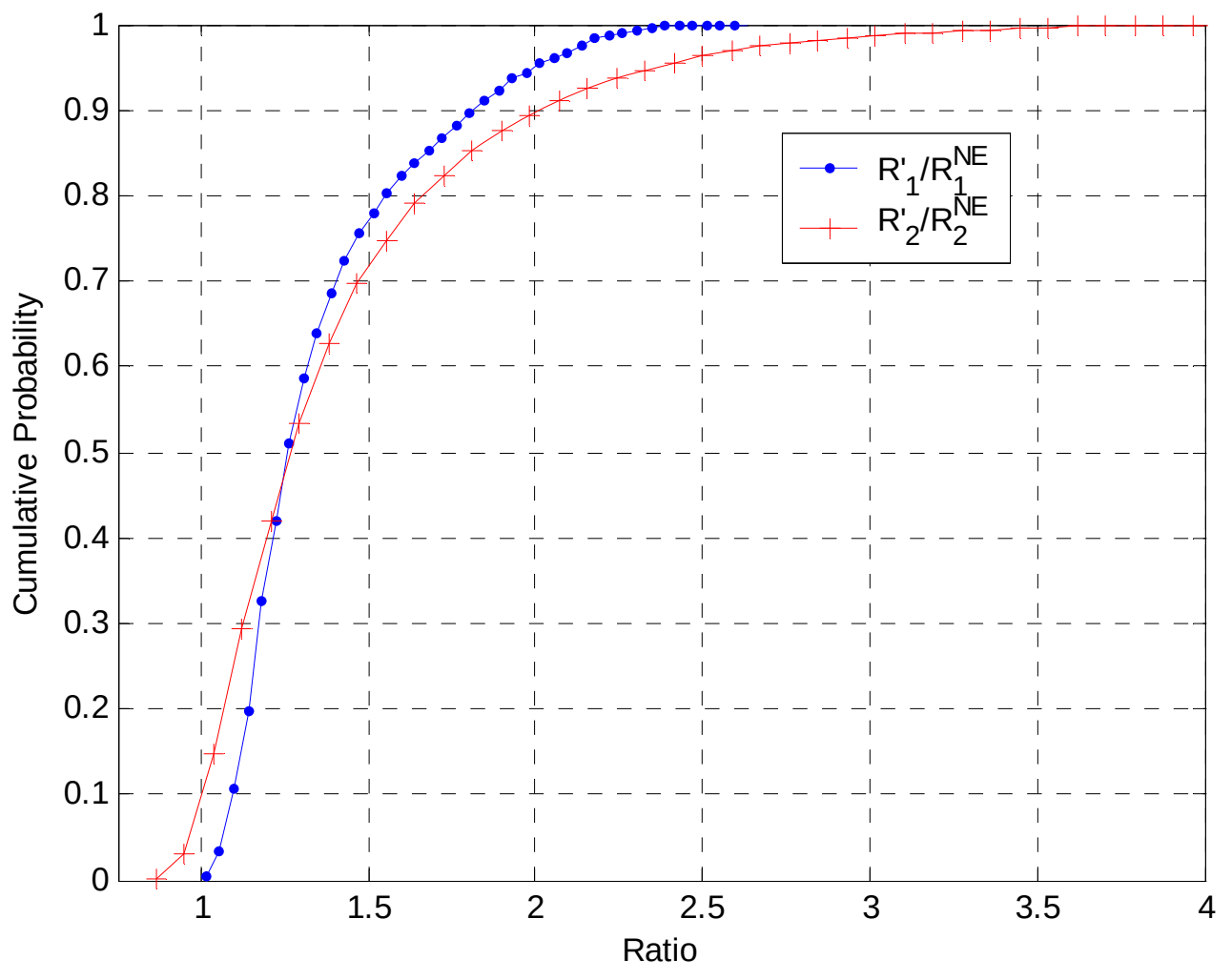

Fig. 8. Cdfs for the ratio of $R_i' / R_i^{NE}$ ($\sum_f \left|H_{12}^f\right|^2 = \sum_f \left|H_{21}^f\right|^2 = 0.5$).

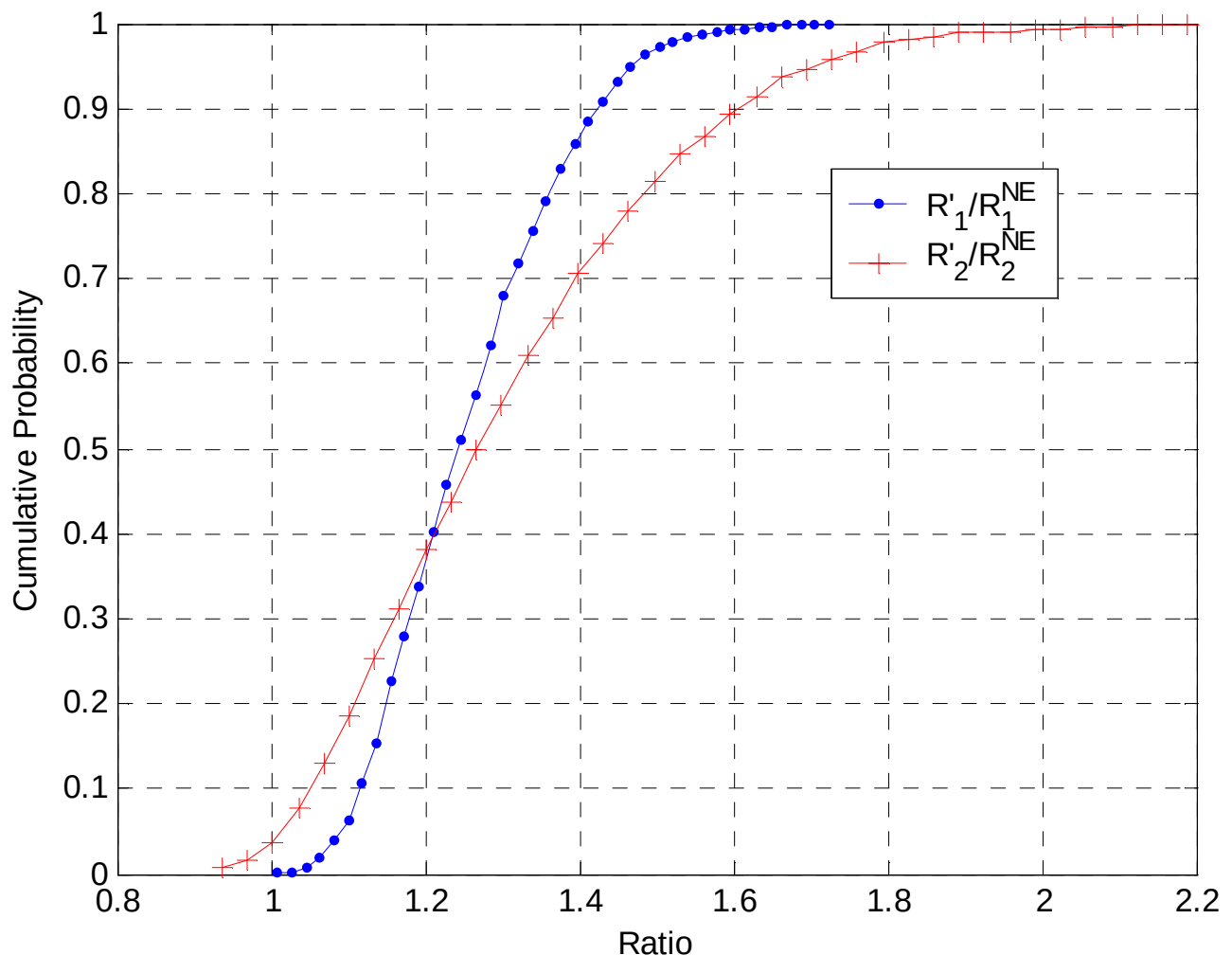

Fig. 9. Cdfs for the ratio of $R_i' / R_i^{NE}$ ($\sum_f \left|H_{12}^f\right|^2 = \sum_f \left|H_{21}^f\right|^2 = 0.25$).

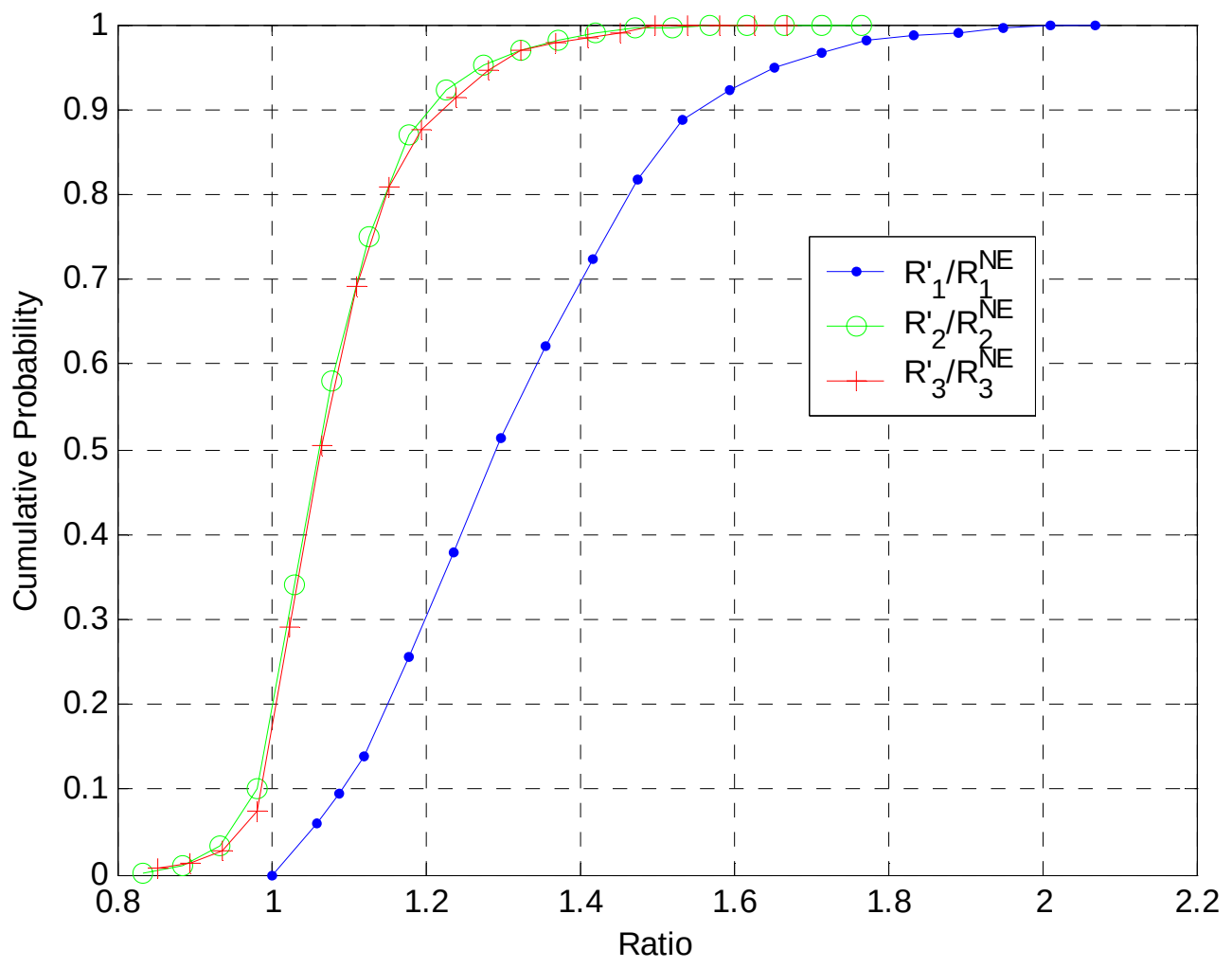

Fig. 10. Cdfs for the ratio of $R_i' / R_i^{NE}$ ($\sum_f \left|H_{ij}^f\right|^2 = 0.25,\ i \neq j$).